\documentclass[aps,prl,amsmath,amssymb,showpacs,twocolumn,groupedaddress,superscriptaddress]{revtex4}

\usepackage[dvips]{graphicx}
\usepackage{bm}
\begin{document}


\title{Asymmetric Reversal in Inhomogeneous Magnetic Heterostructures}

\author{Zhi-Pan Li}
\email[E-mail: ]{zhipan@physics.ucsd.edu} \affiliation{Physics
Department, University of California - San Diego, La Jolla, CA,
92093-0319, USA}
\author{Oleg Petracic}
\affiliation{Physics Department, University of California - San
Diego, La Jolla, CA, 92093-0319, USA} \affiliation{Angewandte
Physik, Universit\"at Duisburg-Essen, 47048 Duisburg, Germany}
\author{Rafael Morales}
\affiliation{Physics Department, University of California - San
Diego, La Jolla, CA, 92093-0319, USA} \affiliation{Departamento de
F\'isica, Universidad de Oviedo, Oviedo 33007, Spain}
\author{Justin Olamit}
\affiliation{Physics Department, University of California, Davis,
CA 95616, USA}
\author{Xavier Batlle}
\affiliation{Physics Department, University of California - San
Diego, La Jolla, CA, 92093-0319, USA} \affiliation{Departament de
F\'isica Fonamental, Universitat de Barcelona, 08028 Barcelona,
Spain}
\author{Kai Liu}
\affiliation{Physics Department, University of California, Davis,
CA 95616, USA}
\author{Ivan K. Schuller}
\affiliation{Physics Department, University of California - San
Diego, La Jolla, CA, 92093-0319, USA}

\date{\today}

\begin{abstract}
Asymmetric magnetization reversal is an unusual phenomenon in
antiferromagnet / ferromagnet (AF/FM) exchange biased bilayers. We
investigated this phenomenon in a simple model system experimentally
and by simulation assuming inhomogeneously distributed interfacial
AF moments. The results suggest that the observed asymmetry
originates from the \textit{intrinsic} broken symmetry of the
system, which results in \textit{local incomplete domain walls}
parallel to the interface in reversal to negative saturation of
the FM. Magneto-optic Kerr effect unambiguously confirms such an
asymmetric reversal and a depth-dependent FM domain wall in
accord with the magnetometry and simulations.
\end{abstract}


\maketitle

Exchange coupling between a ferromagnet (FM) and an antiferromagnet
(AF) has been intensely studied due to the fundamental interest in
inhomogeneous magnetic systems and its central role as a magnetic
reference in various devices. In most magnetic systems, time
reversal symmetry is present and manifested by a
symmetric magnetization curve relative to the origin. This symmetry also requires
that the magnetization reversal from positive to negative
saturation be identical to the reverse process.
However, in a FM/AF system, exchange bias (EB) develops below the AF
N\'eel temperature $T_{N}$ producing a shift ($H_{EB}$) of the
hysteresis loop along the magnetic field axis \cite{01M-B}. Therefore, with the
shift breaking the time reversal symmetry, magnetization reversal symmetry is no longer required. In
fact, asymmetric reversal was observed by polarized neutron reflectometry \cite{18Fitzsimmons},
photoemission electron microscopy \cite{20Krishnan},
magnetotransport \cite{21Kai}, magneto-optical indicator film
\cite{22Nikitenko}, and magneto-optical Kerr effect
\cite{23Eisenmenger}. In some systems the reversal along the
decreasing branch is dominated by transverse magnetic moments which was interpreted as 
due to coherent magnetic rotation. The absence of transverse moments in the increasing branch reversal
was interpreted as domain wall propagation \cite{18Fitzsimmons, 20Krishnan}. Different, even
opposite scenarios were also found \cite{23Eisenmenger, 25Gierlings,
26Leighton}. Despite the well established experimental evidence and
proposed theoretical models \cite{27Li-Zhang,28Beckmann, 34JCamarero}, the origin
of this asymmetry remains a controversial and highly debated issue \cite{34Tillmans}.
This situation is further complicated by the lack of knowledge of the
interface, crystal imperfections, complex FM and AF anisotropy
energies, and training effect. While these factors are important for
each individual system, the fundamental connection of the reversal
asymmetry to the broken symmetry intrinsic in the
inhomogeneous system is overlooked.

In this Letter, we have investigated a simple model system using a
variety of experimental techniques combined with numerical
simulations. We establish a critical link between this unusual
reversal asymmetry with the time reversal asymmetry in these
systems. Namely, in reversal toward the two FM saturated states, the
intrinsic asymmetry gives rise to different competing mechanisms,
thus different reversal processes.

FeF$_2$/(Ni, Py) bilayers were prepared for this study. FeF$_2$ is
an AF with a N\'eel temperature $T_{N}$ = 78~K, and a large uniaxial
anisotropy $K_{u}=1.35\times10^{4}$~kJ/m$^{3}$ along [001]
direction, hence can be considered as a model Ising system
\cite{29Hutchings, Strempfer}, with the AF spins frozen along [001] at low temperatures \cite{30spinflop}. The Ni or
Py (Ni$_{81}$Fe$_{19}$) is polycrystalline with a negligibly small
crystalline anisotropy, except for a small growth-induced uniaxial
anisotropy along FeF$_2$ [001] \cite{31Petracic}. This system is
thus in close approximation with simple theoretical assumptions.

\begin{figure}[htbp]
\includegraphics[width=6.5cm]{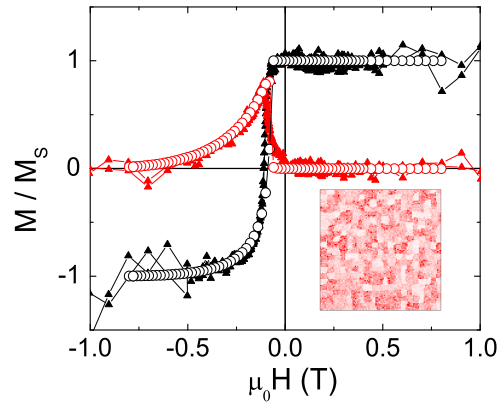}
\caption{(color online) (a) Vector VSM measurement (filled symbol)
and micromagnetic simulation (empty symbol) of
FeF$_2$(50~nm)/Ni(21~nm) at 15~K after field cooling in a 0.2~T
field. Both longitudinal (black symbol) and transverse (red
symbol) components are measured and simulated. The micromagnetic
simulation was performed assuming the FM interfacial layer is
coupled to spatially inhomogeneous uncompensated frozen AF spins,
whose distribution is shown in the inset (500$\times$500 nm$^2$).
The different shades of red refer to the different magnitude of
local uncompensated frozen AF moments with white corresponding to
zero local density.}\label{fig1}
\end{figure}

The bilayer was grown by e-beam evaporation on a single crystal
MgF$_2$(110) substrate, where FeF$_2$ (110) grows epitaxially
untwinned \cite{32Shi-Lederman, 31Petracic}. Vector vibrating sample
magnetometry (VSM) of FeF$_2$ (50~nm) / Ni (21~nm) / Al (7.6~nm)
gives simultaneously the in-plane longitudinal (parallel to the
magnetic field) and transverse (perpendicular to the magnetic field)
magnetic moments \cite{33Olamit, 41Daboo}. The magnetic field is applied
along the FeF$_2$ easy axis [001] with a small misalignment that
defines the sign of the transverse component during reversal
\cite{33Olamit,34Tillmans}. Square hysteresis loops are
found above $T_{N}$ along [001] \cite{31Petracic}. Cooling the
sample in a field $\mu_0H_{FC}$ = 0.2~T from T = 150~K to 15~K
yields an EB field $\mu_0H_{EB}$ = -0.1~T (Fig. \ref{fig1}a) and
virtually no coercivity. Both longitudinal and transverse hysteresis
loops exhibit a clear asymmetry. Starting from positive saturation,
the reversal occurs with a sharp corner in the longitudinal
component and an abrupt increase in transverse component to over
75\% of the saturation magnetization. Then the FM \textit{gradually}
approaches negative saturation, evidenced by the long tail in both
components. A significant non-zero transverse component is found
even at $\mu_0H$ = -0.5~T. In the increasing field sweep, Ni is
saturated almost \textit{immediately} after the reversal. The
asymmetry of the two FM orientations, especially the long tail
around negative saturation, is key to understanding the asymmetric reversal.

We modeled the asymmetric reversal process with micromagnetic
simulations \cite{35oommf} using a 20~nm thick Ni layer with lateral
size 500$\times$500~nm$^{2}$, discretized into
5$\times$5$\times$2~nm$^{3}$ cells. The Hamiltonian $\mathcal{H}$ of
the system is given by,
\begin{displaymath}
\begin{split}
\mathcal{H}=&A\sum_{i\in\{FM\}}[(\nabla\widehat{m}_{ix})^{2}+(\nabla \widehat{m}_{iy})^{2}+(\nabla\widehat{m}_{iz})^{2}]\Delta V \protect\\
&-\sum_{i\in\{FM\}}(K_{u}\widehat{m}_{ix}^{2} \Delta
V+K_{d}\widehat{m}_{iz}^{2}\Delta V+\vec
{H}\cdot \vec m_i)\protect\\
&-J_{FM/AF}\sum_{i\in\{Interface\}}\vec{m}_{i} \cdot
\vec{\sigma}_{i},
\end{split}
\end{displaymath}
where the three summed terms include FM exchange energy, FM
anisotropy and Zeeman energy, and FM/AF interfacial interaction,
respectively. The AF is assumed to be frozen during the hysteresis
cycle, thus its energy contribution remains constant and is not
considered in the Hamiltonian above. $\vec{H}$ is the magnetic field
applied along the $\hat{x}$ axis with $0.5^{\circ}$ misalignment similar to
the experiment. $\vec{m}_i$ and $\Delta V$ are the magnetic moment
and volume of each cell, respectively. The reduced moment
$\widehat{m}_i$ is defined by $\widehat{m}_{i}=\vec m_i / M_S$. We
used the nearest-neighbor exchange constant
$A=3.4$~pJ/m and the saturation magnetization
$M_{S}=494$~kA/m for Ni \cite{36Skomski}. The small
growth-induced anisotropy of the Ni layer is taken into account by a
uniaxial anisotropy along the $\hat{x}$ axis with $K_u=5$~kJ/m$^{3}$
obtained from measurements along the hard axis above $T_N$. The dipolar interaction is
approximated by a shape anisotropy along the $\hat{z}$ axis
(out-of-plane) with $K_{d}=-(\mu_{0}/2)M_{S}^{2}=-150$~kJ/m$^3$,
which keeps the moments in the sample plane and avoids boundary effects.

The AF is modeled by a monolayer of spatially inhomogeneous frozen
moments, $\vec \sigma_i$, exchange coupled to the bottom layer of
the FM with an adjustable interfacial coupling from $J_{FM/AF}\sim
J_{AF}=-0.45$~meV up to $2J_{AF}$ \cite{29Hutchings}. We introduced AF grains of average size
25$\times$25~nm$^{2}$ to simulate the inhomogeneous interfacial coupling\cite{32Shi-Lederman}. 
$\sigma_{i}=-\alpha_{i}S_{i}^{AF}p_{j}$ with $S_{i}^{AF}=2$, consists of
two random quntities: $\alpha_{i}$ denoting the \textit{intergrain} variation, and 
$p_{j}$ the \textit{intragrain} variation. $\alpha_i$ varies as 1$\pm$0.35 between grains, 
while $p_{j}$ varies as (7$\pm$2)\% between cells. This 7\% assumption is based on
recent experiments which found net frozen AF interfacial moments
with about 4\% \cite{06Ohldag} or 7\% \cite{07Kappen} coverage that contribute to EB. 
Crucial parameters for the simulation include the product of the uncompensated moment coverage
and interfacial coupling, and intergrain fluctuation. The former defines the effective coupling
strength. The latter describes the interfacial inhomogeneity modulated over a length scale 
of the grain size (25 nm), comparable with the FM domain wall width 82 nm. 
This spatial modulation of $\sigma_i$ leads to an inhomogeneous pinning on the FM, and is essential to explain
reversal process revealed in the experiment. However, the intensity of the modulation is not essential: 
20\% to 50\% standard deviation in $\alpha_i$ gives similar results. The resultant spatial
variation of $\sigma_i$ is shown in the inset of Figure \ref{fig1}.

\begin{figure}[htbp]
\includegraphics[width=8.5cm]{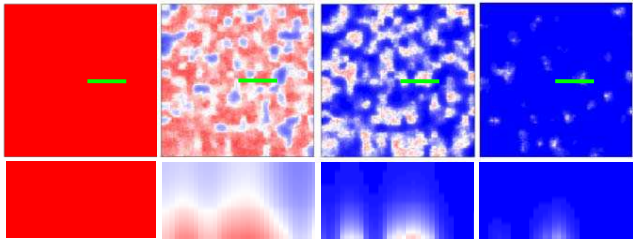}
\caption{(color online) Images in the first row from left to right
are the simulated FM spin configurations (500$\times$500 nm$^2$) at
the FM/AF interface at 0.8, -0.08, -0.36, and -0.8 T, respectively;
the second row shows the simulated FM depth profiles (125$\times$20
nm$^2$), the bottom edge referring to the FM/AF interface) for the
same corresponding field cross sectioned at the green lines. Red,
white and blue corresponds to $M_{x}/M_{s}$ = 1, 0, -1,
respectively, with $x$ being the magnetic field direction.}
\label{fig2}
\end{figure}

The simulation (Fig. \ref{fig1}) matches well both the
longitudinal and transverse hysteresis loops exhibiting the same
asymmetry as the experiment \cite{37coupling1}. The bottom and
side view of the FM spin configuration during the hysteresis
(Figure \ref{fig2}) shows domains evolving both laterally and in
the depth. In positive saturation, the FM is uniformly magnetized
throughout the thickness because both the applied magnetic field
and interfacial coupling favor this orientation. As the magnetic
field decreases, the reversal is initiated from the top of the FM
far away from the interface while the bottom pinned by the AF
remains in the positive direction. An incomplete
(non-$180^{\circ}$) FM domain wall (IDW) is thus formed parallel
to the interface. As the field decreases further, these FM IDWs
slowly shrink laterally and squeeze close to the interface. Even
at $\mu_0H$ = -0.8~T, the FM is not saturated at some interface
regions. This lateral domain formation is the result of the
spatially varying $\sigma_i$. The regions in the FM most resistant
to reversal are where the strongest local interfacial pinning is
found. As the field increases, these regions become nucleation
sites for the development of \textit{local} IDWs both laterally
and in the depth. Therefore, these local IDWs result from the
competition between inhomogeneously distributed interfacial
pinning and the magnetic field. Due to the unidirectional nature
of the AF pinning field, it only competes with the Zeeman energy
in approaching negative saturation, while they both stabilize the
FM when positively saturated. This simulation demonstrates that
the local development of IDWs constitutes the dominant asymmetric
reversal mode. Although similar exchange spring is claimed in
hard/soft magnetic structures \cite{35kneller, 36fullerton}, it does not lead
to asymmetric reversal \cite{Davies}. In addition, this incomplete domain wall is
unusual in EB because the interfacial coupling energy is much weaker
than that in a conventional exchange spring, thus it was never convincingly observed and
was overlooked in most EB studies. 

When a finite anisotropy of pinned AF moments is included in the simulation,
the IDW is pushed into the AF forming a hybrid domain wall across the interface, 
but the main features of the reversal process remain unchanged. Since the 
anisotropy of the FM is usually much smaller than that of the AF,
the FM-side of domain wall dominates the reversal.

\begin{figure}[htbp]
\includegraphics[width=6cm]{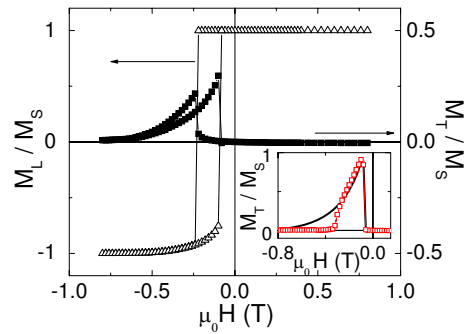}
\caption{(color online) Simulation of the longitudinal (open
symbols) and transverse (filled symbols) hysteresis loop
considering 50~ kJ/m$^3$ in-plane uniaxial anisotropy and
$0.5^{\circ}$ fanning of AF uncompensated moment orientation.
(Inset) Simulated transverse hysteresis loops with uniform (red
open symbols) and inhomogeneous (black line) interfacial
coupling.}\label{fig3}
\end{figure}

The result implies several important features of the local IDW
reversal process. First, the FM domain wall depth-dependence is
crucial for the asymmetric reversal process. An important
signature of this behavior is the asymmetric development of
transverse magnetic moments. This behavior tends to be smeared out
by AF twinning or polycrystallinity, and/or more complicated FM or
AF anisotropy energy terms. It is worth noting that this asymmetry
of approaching two saturated states may seem different from the
asymmetry of two field sweep branches observed before in other
systems, where a sharp corner is found in the decreasing branch
and a rounded one for the increasing one \cite{21Kai,
23Eisenmenger}. However, they are essentially the same except for
the small FM uniaxial anisotropy, thus negligible coercivity in
our system. If the FM uniaxial anisotropy is increased to $K_u$ =
50~kJ/m$^3$ and a $0.5^{\circ}$ fanning of the AF pinning moment
in the sample plane is included, the simulated hysteresis loop
displays the same asymmetry as observed before together with an
irreversible transverse loop (Fig. \ref{fig3}) \cite{21Kai,
23Eisenmenger}. Second, the \emph{local} nature of the IDW due to
the interfacial inhomogeneity is crucial in the model. It leads to
asymmetric lateral domains due to unsynchronized winding of DW in the depth,
and may clarify the present confusion and debate
based on lateral multi-domain observations. It also explains the long
tail of the hysteresis loops, which would otherwise disappear if
$\alpha_i$ is not included as in Kiwi's model (Fig. \ref{fig3}
inset) \cite{14Kiwi}. Since a square hysteresis loop is observed
above $T_N$, this low temperature behavior must arise from the
interfacial inhomogeneity.

\begin{figure}[htbp]
\includegraphics[height=6cm]{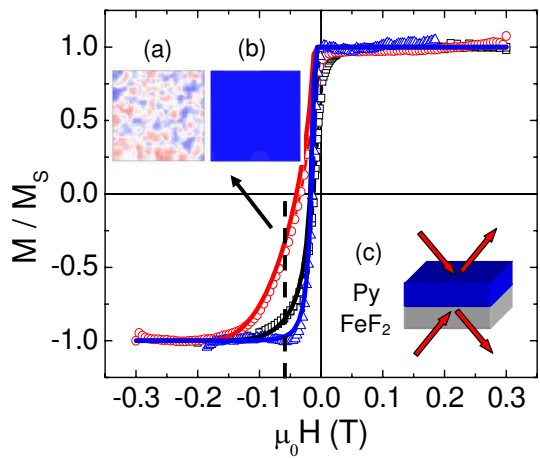}
\caption{(color online) Experiment (open symbol) and micromagnetic
simulation (solid line) on FeF$_2$ (70~nm) / Py (70~nm) at 10~K
after field cooling in a 0.02~T field. Experimental curves
obtained from MOKE measurement from the FM-air (blue triangle) and
FM-AF (red circle) interface and SQUID magnetometry (black
square). The schematic of the MOKE experiment is shown in inset
(c). The upper-left inset shows the simulated FM spin
configuration (500$\times$500 nm$^2$) at the FM-AF (a) and FM-air
(b) interface at $\mu_0H=-0.06$~T.}\label{fig4}
\end{figure}

So far we demonstrated that the local IDWs nucleated in
approaching the negative saturation cause the asymmetric reversal.
This result is unambiguously confirmed by MOKE experiments probing
the FM-air and FM-AF interfaces independently. In this experiment,
a sample with MgF$_2$ (110) / FeF$_2$ (50~nm) / Py (70~nm) / Al
(4~nm) is cooled below $T_{N}$ in $\mu_0$~$H_{FC}$ = 0.02~T, and
MOKE is performed on both the top and bottom surfaces of the
sample with HeNe laser ($\lambda$ = 632.8~nm) at 45 degree
incidence (Fig. \ref{fig4} inset (c)). Probing the depth
dependence of the FM domain structure is possible because the 28
nm penetration depth of the light \cite{CRC} is less than half of
the Py thickness, and both MgF$_2$ and FeF$_2$ are transparent. A
clear difference is seen between the two MOKE measurements (Fig.
\ref{fig4}). Probing the FM-AF interface shows a much more rounded
and longer tail compared with the one from FM-air interface,
confirming the existence of domain structures in the depth. The
sample was also measured using SQUID magnetometry to which the
entire sample contributes equally. The resultant hysteresis loop
lies between the two MOKE loops.

We also performed micromagnetic simulations under identical
assumptions using the same parameters as above to generate the
random frozen AF moments \cite{40Py}. The exponential decay of
MOKE in the FM is simulated by giving each FM discretization layer
in the depth an appropriate weight according to the Py 28~nm
penetration depth. A very good agreement is obtained for all three
hysteresis loops simultaneously with a slight adjustment of the
interfacial coupling \cite{38coupling2}. At $\mu_0H$ = -0.06~T, a
large difference between the two MOKE measurements is observed.
The simulated spin configuration at this field shows that the FM
close to the FM-AF interface is only partially reversed forming
lateral domain patterns, while at the FM-air interface the FM is
fully reversed (inset (a) and (b) of Figure \ref{fig4}). This
confirms that the local IDW model leads to asymmetrically rounded
hysteresis loops.

In summary, we found strongly asymmetric hysteresis loops in a
simple model exchange bias system FeF$_2$/(Ni, Py). By combining
vector magnetometry, MOKE with micromagnetic simulation, we
clearly showed that the asymmetric reversal directly results from
the FM domain structure in the depth due to the broken symmetry at
the interface. The hotly debated issue over the asymmetric
reversal process over the past 5 years solely focused on
lateral FM domains, and its origin was controversial until now.
FM parallel domains were predicted \cite{14Kiwi, 15Stiles}. However, they were
not confirmed experimentally. They were mostly ignored in
microscopy studies \cite{20Krishnan} and
simulations generally assuming the FM to be a single moment \cite{34JCamarero}
or one monolayer \cite{28Beckmann}. This situation was mostly due to
the weak coupling at the FM/AF interface, and limitations of different
experimental and modeling techniques. Dispersions in AF crystallinity
and anisotropy also smear out manifestations of parallel domain walls. 
Our study of a simple EB model system, combining different
experimental and simulation techniques, unambiguously demonstrates the presence of
such domains and their dominant role on the asymmetric reversal.

\begin{acknowledgments}
We thank M. Kiwi, H. Suhl and C. Miller for illuminating
discussions. Work supported at UCSD by US-DOE, at UCD by UC CLE,
Cal(IT)$^{2}$ (Z.-P. L.), Alexander-von-Humboldt Foundation (O. P.),
Spanish MECD (R. M., X. B.), Fulbright Commission (R. M.),
NEAT-IGERT (J.O.), Catalan DURSI (X. B.) and Alfred P.
Sloan Foundation (K. L.).
\end{acknowledgments}


\begin{thebibliography}{32}
\bibitem[{30s()}]{30spinflop}
\bibinfo{note}{Spin-flop is not observed in the samples discussed here.}
\bibitem[{37c()}]{37coupling1}
\bibinfo{note}{The interfacial coupling that fits best to the
  experiment is $J_{FM/AF}$ = -0.88~eV, comparable with the exchange coupling
  in FeF$_2$ -0.45~eV from Ref. 13. This number is not meaningful in its
  absolute value because it depends on the assumed 7\% frozen AF uncompensated moments.}
\bibitem[{40P()}]{40Py}
\bibinfo{note}{Micromagnetic parameters $M_S=800$~kA/m and $A=13$~pJ/m of Py
  used are from Ref. 20.}
\bibitem[{38c()}]{38coupling2}
\bibinfo{note}{The interfacial coupling is found to be -0.75~eV.}
\end{thebibliography}

\end{document}